\documentclass[twoside,fleqn]{article}
\usepackage{espcrc2}
\usepackage{times,mathptm}
\usepackage[dvips]{graphicx}
\newcommand{\fcaption}[1]{\vspace*{-1.1cm}\caption{#1}\vspace*{-0.1cm}}
\newcommand{\Fig}[1]{Fig.~\ref{#1}}

\title{Topological Susceptibility of Monte-Carlo Generated Projected Vortices%
\thanks{Presented by Manfried Faber. Supported in part by 
Fonds zur F\"orderung der Wissenschaftlichen Forschung 
P13997-TPH (R.B. and M.F.) and by DFG under grant En 415/1-1.}}
\author{%
Roman Bertle\address[AIK]{Atominstitut der \"osterreichischen Universit\"aten,
Techn.\ Univ.\ Wien, A--1040 Vienna, Austria}, 
Michael Engelhardt\address{Institut f\"ur Theoretische Physik,
Universit\"at T\"ubingen, Auf der Morgenstelle 14, 72076 T\"ubingen, Germany}
and Manfried Faber\addressmark[AIK]}
\begin{document}
%
%
\begin{abstract}
We determine the topological susceptibility from center projected vortices
and demonstrate that the topological properties of the $SU(2)$ Yang-Mills
vacuum can be extracted from the vortex content. We eliminate spurious 
ultraviolet fluctuations by two different smoothing procedures. The 
extracted susceptibility is comparable to that obtained from full field 
configurations.
\end{abstract}
\maketitle
One of the main aims of present day lattice investigations is to obtain
a consistent picture of the QCD vacuum. During the last years, the center
vortex model turned out to be a good candidate for that. The vortex model 
was invented at the end of the seventies \cite{vort78}. Due to the lack of 
an identification method for vortices, almost no numerical investigations 
were done for 25 years. Maximal center gauge \cite{deb97} and center 
projection gave us a means to identify vortices and led to new 
investigations about the predictive power of the vortex model.
The vortex model explains the confinement properties of the QCD vacuum 
\cite{vort78,deb97,kt98,fgo98,tlang}. By the idea that monopoles are
located on vortices \cite{dfgo98}, it is even strongly related to the dual 
superconductor picture of confinement. As recently \cite{er00} suggested, 
the vortex picture is also related to the topological properties of the 
QCD vacuum. We report on numerical calculations which support this statement.

We extract center vortices from lattice configurations by direct maximal 
center gauge \cite{dfggo98}, i.e.~by a maximization of the gauge fixing 
functional
\begin{equation}
\max_{G} \sum_{i} \left| \mbox{tr} \, U_i^G \right|^{2}
\label{gfunct}
\end{equation}
under gauge transformations $G$, where the $U_i $ are the link variables.
This functional biases link variables toward the center of the gauge 
group, e.g.~in the $SU(2)$ case considered henceforth, toward the elements 
$\pm 1$. Physically, the idea is to transform as much physical information
as possible to the center part of the configuration. In a second step,
center projection, the magnetic flux is concentrated in tubes (P-vortices)
which propagate in time and form closed two-dimensional surfaces in dual
space. Since the Pontryagin index $Q$, a four-dimensional integral over
the scalar product of electric and magnetic field strengths, is a
topological quantity, one can hope that it is effectively unchanged by a
compression of electric and magnetic fluxes.

For thin center vortices in the continuum, the topological charge $Q$ is
given by their self-intersection number \cite{er00}
\begin{equation}\label{yidres}
Q = -\frac{1}{16} \epsilon_{\mu \nu \alpha \beta }
\int_{S} d^2 \sigma_{\alpha \beta }
\int_{S} d^2 \sigma^{\prime }_{\mu \nu }
\delta^{4} (\bar{x} (\sigma ) - \bar{x} (\sigma^{\prime } ) ),
\end{equation}
where $\bar{x} (\sigma )$ parametrizes the two-dimensional vortex surface 
$S$ in four-dimensional space-time. This expression is only well defined if 
the orientation of $S$ is specified. Lattice calculations show 
that, in the QCD vacuum, projected vortices have a random structure and are 
in general non-orientable \cite{bfgo99}. This demonstrates that vortex 
surfaces consist of patches of alternating orientation which are separated 
by closed lines where the orientation switches. These flips of orientation 
of P-vortices are related to the color structure of the original thick 
vortices. In an Abelian description, these lines are world-lines of Abelian 
monopoles with the two halves of the magnetic monopole flux going in 
opposite directions. The positions of the monopoles depend on the color 
structure of the vortices and the choice of the U(1) subgroup.

\begin{figure}
\centering
\includegraphics[width=0.45\linewidth]{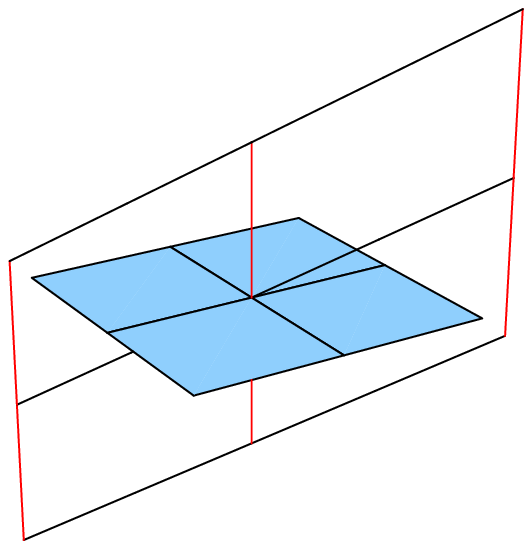}
\vspace{-4mm}
\fcaption{Two orthogonal sets of 4 plaquettes each at an intersection 
point in four-dimensional space.}
\label{4x4}
\end{figure}
According to Eq.~(\ref{yidres}), in the continuum an intersection point 
contributes $\pm 1/2$ to $Q$. On the lattice, an intersection point joins 
$4\times 4$ plaquettes, see \Fig{4x4}; therefore, every joined pair of 
electric and magnetic plaquettes contributes $\pm 1/32$. There are further 
contributions to $Q$ from  ``writhing'' points. These are lattice sites 
where electric and magnetic plaquettes share a point of the four-dimensional
lattice and belong to the same surface region. In spite of these 
contributions, $Q$ remains quantized.

\begin{figure}
\centering
\includegraphics[width=0.45\textwidth]{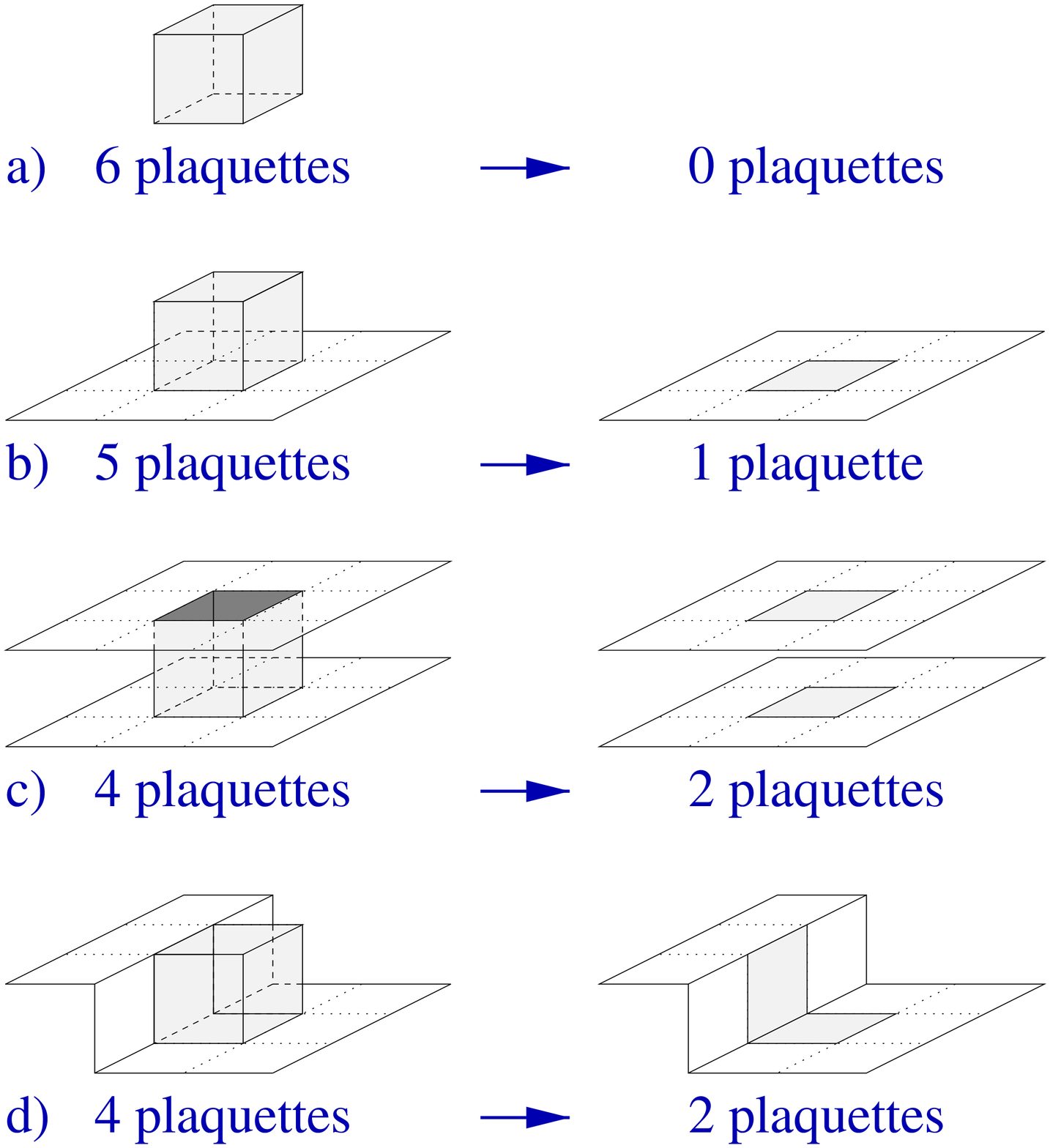}
\vspace{4mm}
\fcaption{Different elementary cube transformations.}
\label{steps}
\end{figure}
Using the above prescription, we will determine $Q$ for Monte-Carlo 
generated gauge field configurations. However, first we have to solve 
three problems. Due to their transverse fluctuations inside the original 
thick vortices, P-vortices are plagued by spurious ultraviolet fluctuations.
We remove them by the elementary cube transformations (a)-(d) in \Fig{steps}
\cite{bfgo99} or by blocking the gauge field configuration such as
to arrive at a two or three times coarser lattice. A further
problem is that Abelian monopole lines determined by Abelian projection are
not always on vortices; some $3\%$ of monopole cubes are not pierced by
P-vortices \cite{dfgo98}. Instead of trying to adjust monopole trajectories
to vortices, we randomly assign orientations to the vortex plaquettes, with
two different choices of bias allowing us to explore the extreme cases of
either maximizing or minimizing the monopole line density. In our 
measurements, the monopole line density varies between the two extremes by
a factor of around ten. The third problem is that, in contrast to the 
continuum, on the lattice P-vortices in general do not intersect at points,
they intersect along lines, and some monopole lines coincide with vortex
intersections and writhing points. We remove these artefacts by transfering
the P-vortices to finer lattices of $1/3$ or, if necessary, $1/9$ the 
lattice spacing. Then we apply elementary cube transformations once at each
lattice site whenever this allows us to remove an instance of the above 
coarse graining problems.

\begin{figure}[!b]
\centering
\includegraphics[width=0.45\textwidth]{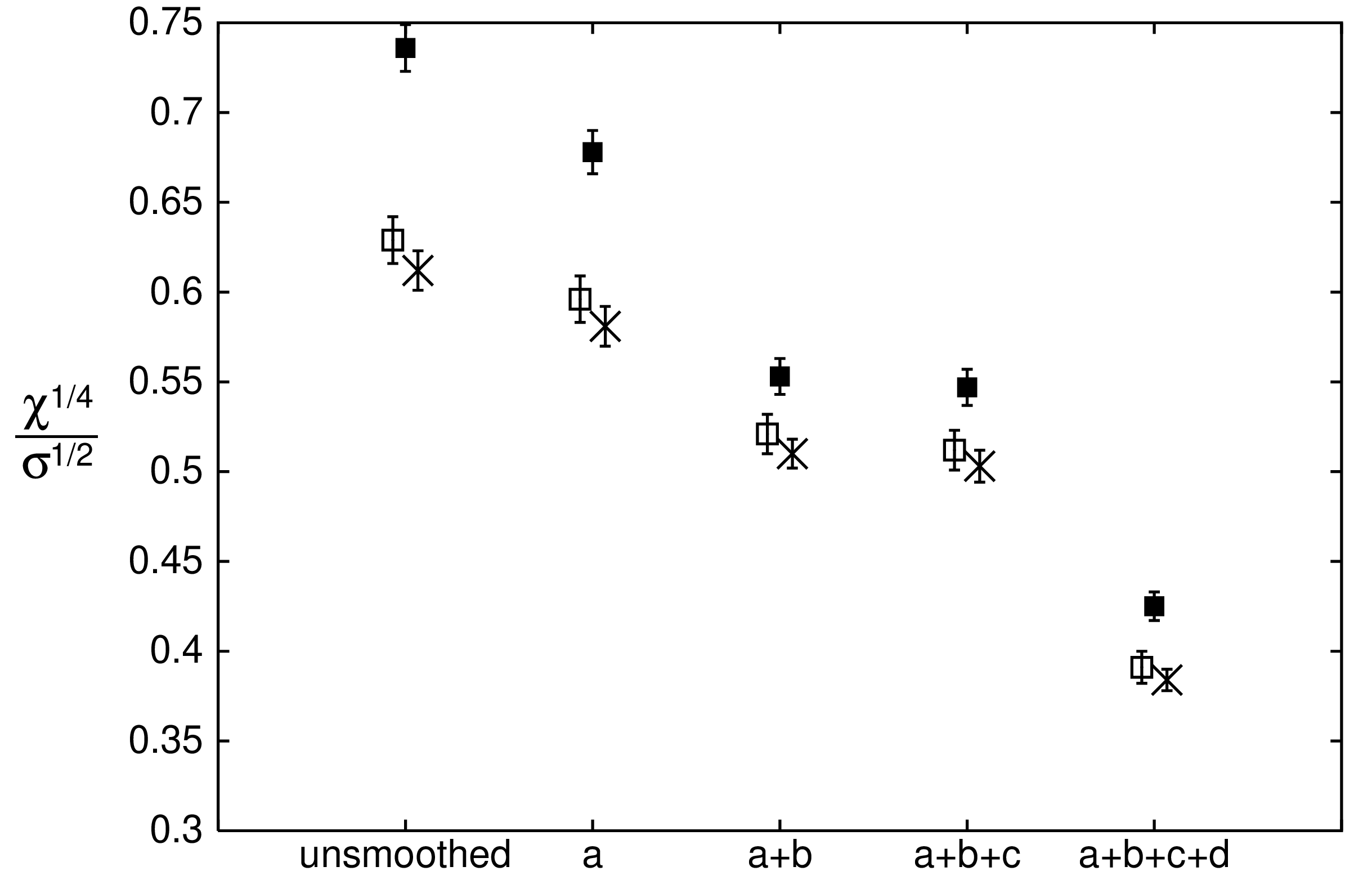}
\includegraphics[width=0.45\textwidth]{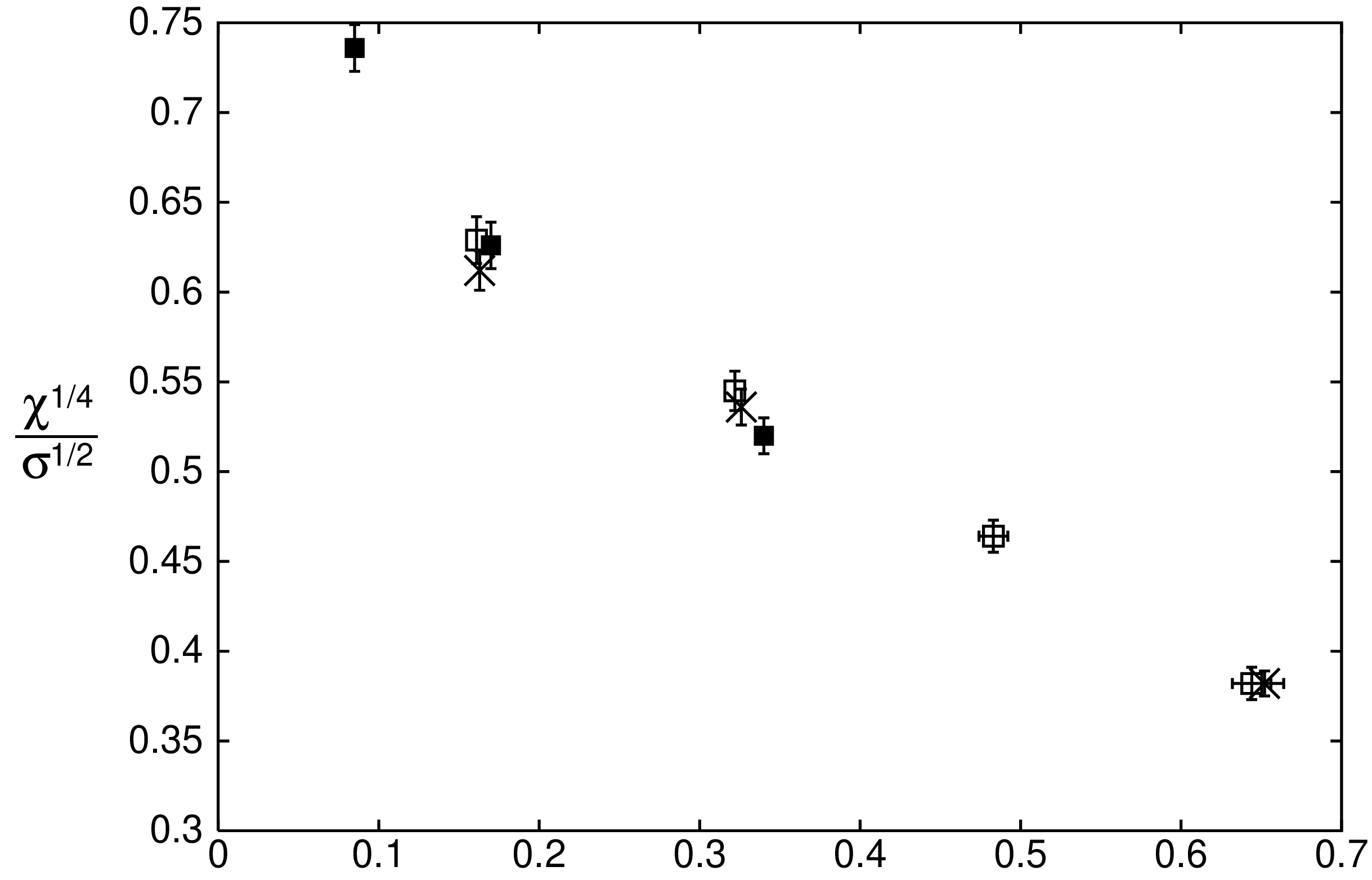}
\vspace{4mm} \fcaption{Fourth root of the topological susceptibility
$\chi $ carried by center projection vortices, in units of the square
root of the string tension $\sigma $, simultaneously extracted from the
center projected lattice ensemble. Values are shown as a
function of the smoothing steps (top) and of the blocking
scale a/fm (bottom).}
\label{datsusc}
\end{figure}
\Fig{datsusc} depicts the results of the numerical measurements of the
topological susceptibility $\chi $ for the coupling $\beta =2.5$ on a 
$16^4 $ lattice (filled squares), where $1156$ samples were taken, and also
for $\beta =2.3$ both on a $16^4 $ lattice (crosses, $1183$ samples) and on
a $12^4 $ lattice (open squares, $4622$ samples). The displayed results were
obtained using the maximal monopole density. Despite the monopole line 
density varying by a factor of around ten when instead using the minimal
monopole density, $\chi$ only differs by at most 1\%. This observation
agrees with the random vortex surface model \cite{rsm00,preptop}. The
vertical error bars in \Fig{datsusc} are compounded from the statistical
uncertainty of the susceptibility and the statistical uncertainty of the
string tension measurements. The latter uncertainty in addition leads to
the horizontal error bars in the bottom panel in \Fig{datsusc}, since the
evaluation of $\sigma a^2 $ was also used to determine the lattice spacing
$a$ by equating $\sqrt{\sigma } =440$ MeV. Since smoothing step (d),
cf.~\Fig{steps}, leads already to a lowering of the string tension
\cite{bfgo99}, we expect the physical $\chi$ between the (a)-(c) and (a)-(d)
values:
\begin{equation}
(166\, \mbox{MeV} )^4 \leq \chi_{phys} \leq (230\, \mbox{MeV} )^4
\end{equation}
in the upper part of \Fig{datsusc}. The lattice spacing which should be
reached by blocking is related to the thickness of the physical vortices
and roughly lies between $0.4$ fm and $0.6$ fm. From the bottom panel
in \Fig{datsusc} we obtain therefore the estimate: 
\begin{equation}
(174\, \mbox{MeV} )^4 \leq \chi_{phys} \leq (224\, \mbox{MeV} )^4.
\end{equation}
These ranges for the topological susceptibility correspond well with values
extracted from the full $SU(2)$ lattice Yang-Mills ensemble.

It is interesting to consider the contribution of the intersection points
alone to the topological susceptibility \cite{protop}. It is suppressed by
a factor of around $2^4$. Therefore, the value of $\chi$ is dominated by
the contributions from the writhing points. This result agrees with the
invariance of $\chi$ under strong changes of the monopole configurations,
which can only influence the contributions from intersection 
points \cite{protop}. In addition, even the contribution of the intersection
points alone varies only by about 5\% with the monopole density. Thus, it 
seems that the minimal monopole density required by the non-orientability of
the vortices is sufficient to randomize the relative surface orientations at
the intersection points.

The results which we obtain for the topological susceptibility support the
vortex picture of the QCD vacuum, i.e.~thick random vortices with color
structure. These vortices can be located in an appropriate gauge like the
maximal center gauge. Center projection of the field configurations
corresponds to a compression of the magnetic flux into quantized tubes.
Thick vortices can explain the string tension between color charges and
its Casimir scaling properties, and, as shown within this work,
the topological properties of the QCD vacuum.
%
%
%
%

%

\end{document}